\documentstyle[12pt,preprint,pre,aps,eqsecnum,epsf]{revtex}

\tightenlines
 
\begin{document}
 
\draft
 
\title{Model for Anisotropic Directed Percolation}
\author{V. Lien Nguyen}
\address{Institute of Physics, P.O. Box 429, Bo Ho, Hanoi 1000 Vietnam}
\author{Enrique Canessa\thanks{E-mail: canessae@ictp.trieste.it}
}
\address{
International Centre for Theoretical Physics, P.O. Box 586,
34100 Trieste, Italy}
\date{\today}
 
\maketitle
 
\begin{abstract}
\baselineskip=22pt

We propose a simulation model to study the properties of directed
percolation in two-dimensional (2D) anisotropic random media.
The degree of anisotropy in the model is given by the ratio $\mu$
between the axes of a semi-ellipse enclosing the bonds that promote
percolation in one direction.  At percolation, this simple model
shows that the average number of bonds per site in 2D is an invariant
equal to 2.8 independently of $\mu$.  This result suggests that
Sinai's theorem proposed originally for isotropic percolation is also
valid for anisotropic directed percolation problems.  The new invariant
also yields a constant fractal dimension $D_{f} \sim 1.71$ for
all $\mu$, which is the same value found in isotropic directed
percolation ({\it i.e.}, $\mu = 1$).  
\end{abstract}
 

\baselineskip=22pt

\section{Introduction}

Critical phenomena \cite{Stau79} in anisotropic systems without
equivalent nearest neighbors constitute an interesting research
topic \cite{Enti77}.  A universal formula for percolation thresholds,
that involves the dimension of the anisotropic lattice and an
arithmetic average of the coordination number for different anisotropic
lattices, has been recently postulated in Ref.\cite{Gala97}.  The
extension of these studies to more complex problems, such as directed
percolation (DP), and more complex systems, such as anisotropic random
systems, is yet to be addressed.  In this context, random systems are
good candidates to model anisotropy since they do not have equivalent
nearest neighbors nor equivalent sites at all lengths.

In this work we propose a simple simulation model to study the
properties of DP in two-dimensional (2D) anisotropic
random media.  The degree of anisotropy is computed by means of
the ratio $\mu = y_{c}/x_{c}$ between the axes of a semi-ellipse
enclosing the bonds that promote percolation in one direction, such
that $y \le x$ (see Fig.1).  As a function of the order parameter
$\mu$ and at the percolation threshold, we measure the
correlation length exponent $\nu$ and the fractal dimension $D_{f}$
of the largest percolating clusters (in systems of up to 51200
random sites).  In the present model, the well-known scaling
exponents of isotropic DP follow by simply setting $\mu = 1$.

At percolation threshold, our model shows that the average number
of bonds per site for DP in anisotropic 2D random systems is an
invariant ($B_{c}=2.8$) independently of $\mu$.  This result
suggests that the Sinai theorem, proposed originally for isotropic
percolation (IP), is also valid for anisotropic DP problems.  The new
invariant also yields a constant $D_{f} \sim 1.71$ for all
$\mu$, which corresponds to the value of isotropic DP.

The paper is organized as follows.  In the next section we outline our
model.  In Sec.III, we present the results of our simulations and
discuss the effects of $\mu$ on the scaling exponents.

\section{The Model}

In order to simulate DP in 2D anisotropic random media we develop
a simulation algorithm similar to the one used in Ref.\cite{Nguy95}.
The coordinates of $N$ sites are generated at random in a square box
of size $L = N^{1/2}$.  The simulation length unit is chosen such that
the density of sites, namely $n$, in the box is always unity regardless
of the total number of sites $N$. The percolation is then checked over
sites from the left edge towards the right edge of the simulation box
({\em i.e.}, along the x-axis in Fig.1).  A periodical
boundary condition is applied in the vertical $y$-direction.

In Fig.1 we show a `particle' that moves from $i$ to $j$. The moving is
allowed whenever the site $j$ is contained within the shaded elliptical
area. In our simulations, the degree of anisotropy is given by the 
parameter $\mu = x_{m}/y_{m}$, where $x_{m}$ is the longer and $y_{m}$
is the shorter axis of a semi-ellipse, {\it i.e.}, $\mu$ is the ratio
of the maximum `hopping distances' along the $x$- and $y$-axes.

In the standard 2D isotropic DP there are three possible equivalent
directions to move: up, down and forward.  This situation in our model
is attained by setting $\mu$ = 1.  In the limit $\mu \rightarrow \infty $,
the model tends to the one-dimensional (1D) percolation problem.  Thus,
simulation results using the present 2D percolation model will reveal
features of the crossover from the standard (say, isotropic) DP to the
1D percolation problem.  For intermediate values of $\mu >1$ our model
features anisotropic DP.

For a given value of the anisotropy parameter $\mu$ and for a given
realization of random site coordinates $\{ x_{i}, y_{i} \}$, in a
sample of size $N = L \times L$, we study percolation from the left-
to the right- simulation box edge.  At the percolation
threshold, we obtain the critical value of the semi-ellipse-axis $y_{m}$:
$y_{c}(N) = y_{m}(N)$ and the mass $M$ of the critical cluster:
$M(N)$ = "{\sl total number of sites belonging to the largest cluster
at percolation}". These quantities, are then averaged over a great
number $K$ of random realizations of site coordinates for the same
sample size $N$ which result on the average quantities
$Y_{c}(N) = <y_{c}(N)>$ and ${\cal M}(N) = <M(N)>$, respectively.

In general, the dependence of the averages $Y_{c}(N)$ and ${\cal M}(N)$
on the samples size $N$ is a consequence of the finite size effects of
the percolation problem.  In order to quantify these effects,
the present simulations were performed at different
$N$ = 400, 800, 1600, 3200, 6400, 12800, 25600 and 51200. 
Accordingly, the number $K$ decreases from $10^{4}$ to $10^{2}$
such that the product of the numbers $K \times N$ is
approximately the same for all sample sizes $N$ in our study.

Along with these average quantities, we also calculate
the moments
\begin{eqnarray}
\delta y_{c}(N) & = & < [y_{c}(N) - Y_{c}(N)]^{2} >^{1/2} \;\;\; ,
                           \;\;\; \label{eq:delta} \\
\delta M(N) & = & < [M(N) - {\cal M}(N)]^{2} >^{1/2} \;\;\; ,
\end{eqnarray}
and also the next-order moments, which are used to estimate
the statistical errors of our simulation results. 

The present measurements are performed for various values of 
$\mu = 1, 2, 3, 4, 5, 6, 7$ and $8$.
As can be seen from the results discussed in the next section, the
greater the value of $\mu$, the stronger the finite size effects are.
We verify that for $\mu =8$ simulations can only been carried out
in samples of size $N \ge 3200$.

Following the well-known finite-size scaling procedure suggested in
Ref.\cite{Levi75}, the critical exponent $\nu$ of the percolation
problem is defined from the scaling expression
\begin{equation}\label{eq:mu}
   \delta y_{c}(N) \propto L^{-1/\nu } \;\;\; ,
\end{equation}
where $\delta y_{c}(N)$ is given in Eq.(\ref{eq:delta}).
Note that in the present study percolation is checked 
by the longitudinal direction only (the $x$-axes in Fig.1), then
the exponent $\nu$ in Eq.(\ref{eq:mu}) should be identified with
the parallel $\nu_{||}$ (see \cite{Stau79}).

\section{Results and Discussion}

In Fig.2(a) the quantities $-\ln \delta y_{c}(N)$ are plotted versus
$\ln L \equiv \ln N^{1/2}$ for different values of the order
parameter $\mu$.  The slopes of the fitting lines give the
corresponding values for the exponent.  Thus we found that
for the largest $\mu=8$, $\nu=1.14$ and for $\mu =2$
we measured $\nu = 1.48$. Other values are given in the figure.
From these calculated moments and the
linear fitting procedure, we estimate the statistical error to be less
than $0.02$ for all values of $\nu$ shown in this figure.  

Results for the DP limiting case $\mu = 1$ has been
previously reported by one of us \cite{Nguy95}.  In this case, the value
$\nu \approx 1.65 \pm 0.02$ is known as the universal value of $\nu_{||}$
for a whole class of istropic DP models in 2D.
As the amount of anisotropy increases, {\it i.e.} $\mu > 1$, the
correlation length exponent $\nu$ decreases.  Since this decrease is
initially very fast to then become smoothly, it is not possible to
obtain the whole crossover from 2D to 1D directed percolation 
for the behaviour of $\nu$.
That is, the decrease from $\nu \approx 1.65$ in 2D-DP to the
limit $\nu = 1$ in 1D is limited by the size of our simulation box.
The finite-size effects in correspondence to the different values of $\mu$
(and, therefore, to the different degrees of anisotropy in the 2D random
systems) are in fact equivalent to those discussed in great detail
in Refs.\cite{Nguy95,Nguy79} for anisotropic percolation and isotropic DP.

By using the values of $\nu$ in Fig.2(a),
the critical `{\it radius}' $y_{c} = y_{c}( N \rightarrow \infty )$
is determined from the scaling expression
\begin{equation}\label{eq:yc}
| y_{c} - y_{c}(N) | \propto  L^{-1/\nu } \;\;\; .
\end{equation}
In Fig.3(a), (b) and (c) the quantities $y_{c}(N)$ are plotted
versus $L^{-1/\nu }$.  From these plots we obtain $y_{c}$ by taking
the asymptotic values $N\rightarrow \infty$ for all $\mu$ studied.
The estimated values of $y_{c}$ are also shown in this figure.

Very remarkably, our simulations show that for all $\mu$ considered
the quantity $\mu \times y_{c}^{2}(\mu)$ is in fact a constant.
Since $(\pi /2) \mu y_{c}^{2} \equiv (\pi /2) x_{c} y_{c}$ is the
area of the critical semi-ellipse at percolation, then our results
suggests that Sinai's theorem \cite{Sinai}, proposed originally
for IP, is also valid for 2D anisotropic DP problems.
In this respect, we emphasize again that our length unit should be
taken as $n^{-1/2}$ for a system with site concentration $n$. 

Thus, our simulations leads to the new invariance
\begin{equation}\label{eq:inv}
B_{c}^{(d)} = S_{c} n \equiv (\pi /2)n \mu y_{c}^{2} = 2.82 \pm 0.02 \;\;\; ,
\end{equation}
where $n$ is the site concentration ({\it e.g.}, the donor concentration
in doped semiconductors), $S_{c}$ is the area of the critical
semi-ellipse and $B_{c}^{(d)}$ is the mean number of connected
bonds per site at percolation.  The invariance of Eq.(\ref{eq:inv}) may be
somehow related to the fractal behavior of the critical clusters
as we shall discuss below.

Let us determine first the fractal dimension $D_{f}$ of the critical
percolation cluster using a standard procedure based on the scaling
expression \cite{Mand82}
\begin{equation} \label{eq:ml}
{\cal M}(N)_{L \rightarrow \infty} \propto  L^{D_{f}}.
\end{equation}
In Fig.2(b) the quantities $\ln {\cal M}(N)$ are plotted against $\ln L$
for different values of the anisotropy parameter $\mu$.

Very surprisingly we found that the fractal dimensions $D_{f}$, as
determined from the slopes of the fitting lines for various values
of $\mu$ in Fig.2(b), seem indeed to be constant and independent of $\mu$
within our simulation errors. We estimate
${\cal D}_{f}(\mu) \equiv D_{f} \approx 1.71 \pm 0.02$ for all $\mu$,
which corresponds to about the same value of the isotropic DP model
with $\mu = 1$.

At a first glance this result might rise some doubts, but we believe
it can be understood in connection with the invariant given in 
Eq.(\ref{eq:inv}).  The invariance of $B_{c}^{(d)}$, with respect to
changes in the anisotropy parameter $\mu$, implies that the average
number of connected bonds at percolation is independent of $\mu$. 
If we assume the percolation process within an elementary semi-ellipse 
(as in Fig.1) to be the `originating percolation rule', then the
invariance of Eq.(\ref{eq:inv}) could mean that the law to generate
percolation clusters remains unchanged as $\mu$ varies.  If this
conjecture is right, we could suggest here a more general statement
for all types of percolation models which are related to each other 
by the Sinai theorem; in these cases, the fractal dimensions of the
percolation clusters could all be the same.

It should be noted that our simulations are limited to $1 \le \mu \le 8$.
It is in this range that we observed the invariance of $B_{c}$ and the
constant value for $D_{f}$.  We believe that these features are
maintained for a larger range of $\mu$ values.  However, it is not
feasible to increase $\mu$ and the sample size $N$ simultaneously and
get to the point where the present 2D simulation model crosses to 
the 1D case ({\em i.e.}, $\mu \rightarrow \infty$).

To conclude, we have suggested a model for anisotropic directed
percolation (ADP) and have presented the first simulation results
for the main critical exponents of the model
in 2D random systems.  Quite surprisingly, we have found a new invariance
for the average number of connected bonds at percolation due to presence
of a suitable external force ({\em e.g.}, shear stress, magnetic field,
{\em etc}).  Our simulations show that the product $\mu \times y_{c}^{2}$
is a constant for all $\mu$'s considered.  This invariance should be in
close relation to the value of $D_{f}$.

We strongly believe the present model of ADP could be important to
describe some physical phenomena such as hopping conduction in anisotropic
$n$-Ge and $n$-Si under strong electrical fields, where the impurity wave
functions are anisotropic and the conduction band splits into one
ellipsoid \cite{Shkl84}.  Our measurements could be useful, for
instance, in the expressions for the hopping resistivity in 2D anisotropic
random media.  The new invariance $B_{c}^{(d)}\sim 2.8$ could be used in
these systems similarly to the invariance $B_{c}^{(i)}=4.5$ for IP
in a circle problem \cite{Shkl84}.  We hope the present model will
stimulate further investigations on this direction.

\section*{Acknowledgment}
 
One of the authors (N.V.L.) would like to thank the Condensed Matter Group
at ICTP, Trieste, for financial support.

\newpage

\begin{figure}[]
\caption [A Picture]
{\protect\normalsize
The anisotropic directed percolation model. Percolation
from site $i$ to $j$ is allowed whenever the site $j$ is contained
within the shaded elliptical area.  The degree of anisotropy is given
by the ratio $\mu = x_{m}/y_{m}$, where $x_{m}$ is the longer and
$y_{m}$ is the shorter axis of the semi-ellipse. The case $\mu = 1$
yields the standard isotropic DP.  At percolation, the average number
of bonds per site (dots in this figure) is 2.8.
}
\end{figure}

\begin{figure}[]
\caption [A Picture]
{\protect\normalsize
Simulation data and linear fitting for:
(a) $-\ln \delta y_{c}(N)$ of Eq.(\ref{eq:delta}) plotted against
the sample size $\ln L$ for several values of the anisotropy
parameter $\mu$;
(b) $\ln {\cal M}(N)$ of Eq.(\ref{eq:ml}) plotted against the
sample size $\ln L$ for several values of the anisotropy
parameter $\mu$.  Within measurement errors, most fitted lines
are parallel giving the average constant value $D_{f} \sim 1.71 \pm 0.02$.
Note that fluctuations around this value are not systematic as $\mu$
increases.
}
\end{figure}

\begin{figure}[]
\caption [A Picture]
{\protect\normalsize
Simulation data and linear fitting for
$y_{c}(N)$ of Eq.(\ref{eq:yc}) plotted against $L^{-1/\nu }$.
(a) $\mu$=1, 2 and 3; (b) $\mu$=4, 5 and 6; (c) $\mu$=7 and 8.
The threshold $y_{c}$ for each $\mu$ studied is estimated from
the asymptotic values $N\rightarrow \infty$.
}
\end{figure}

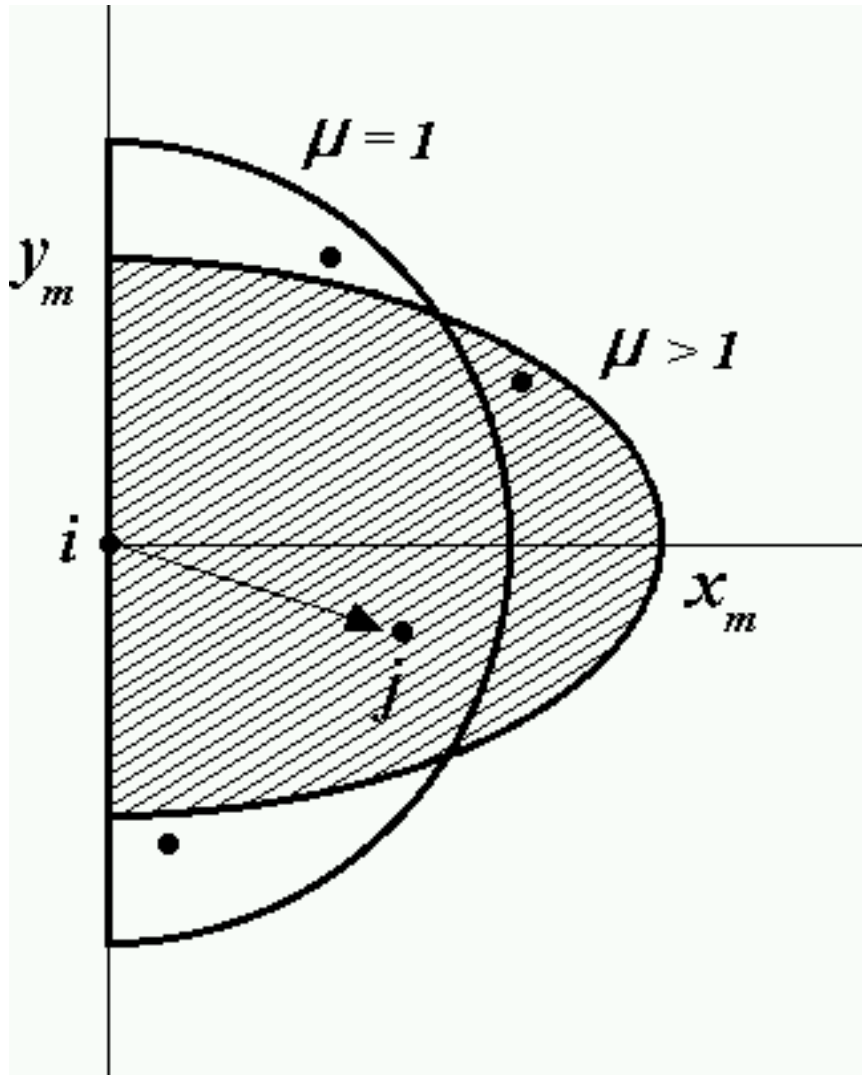
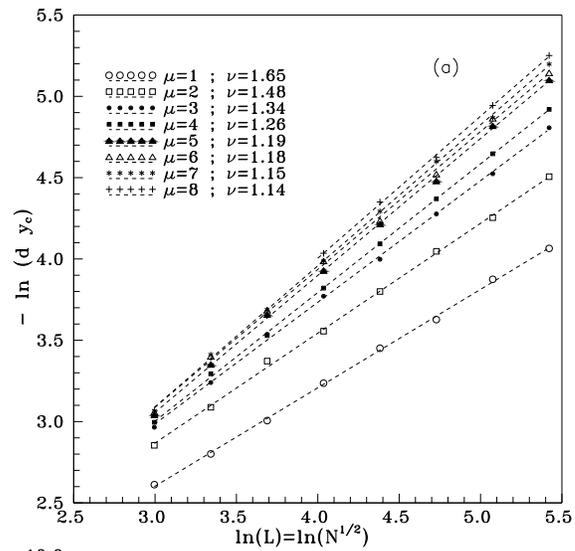
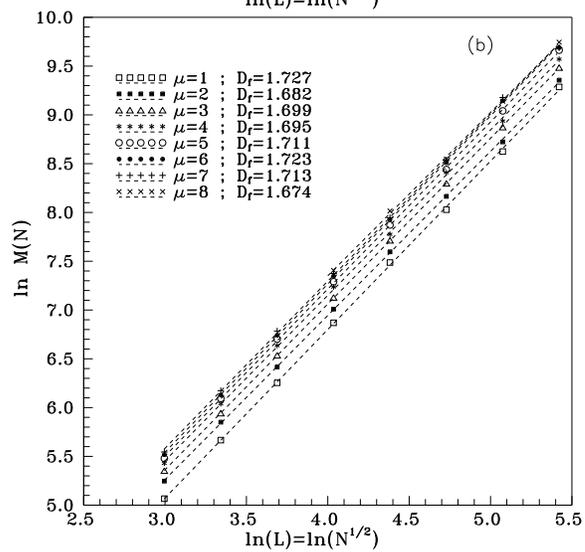
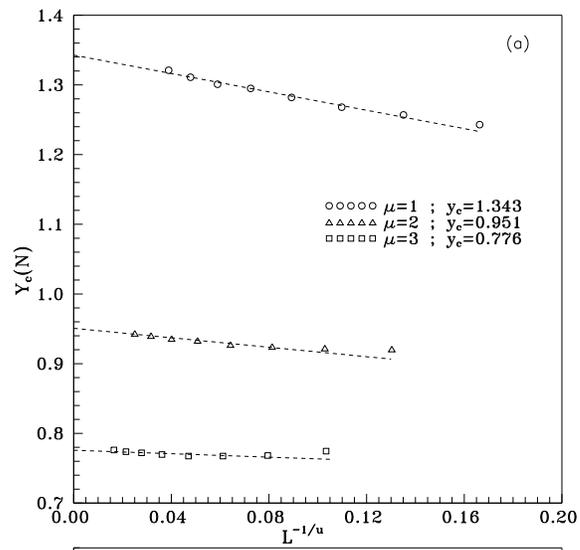
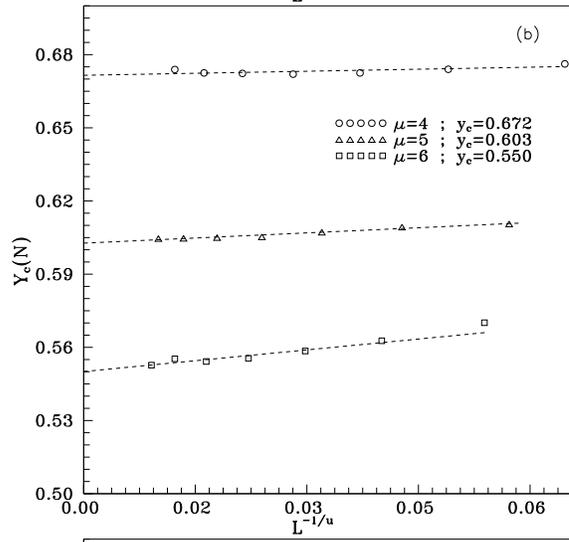
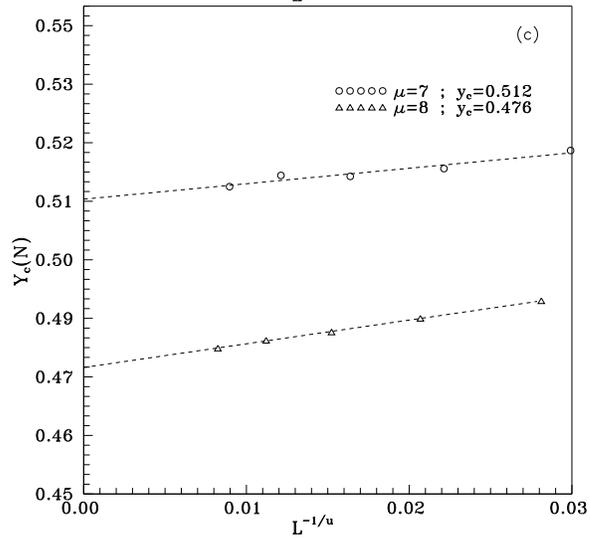
\end{document}